\documentclass[11pt,a4paper]{article}
\pdfoutput=1
\usepackage{pdfpages}
\usepackage{jheppub}
\usepackage{float}
\usepackage{mathrsfs}
\usepackage{amsfonts}
\usepackage{amsmath}
\usepackage[nomessages]{fp}

\def \sd {S^{d_1}\times S^{d_2}}
\def \( {\left(}
\def \) {\right)}
\def \go {g_{(0)}}
\def \gd {g_{(d)}}

\newcommand{\addBHFigure}[4]{
	\begin{figure}[H]
	\centering
	\includegraphics[width=.48\textwidth]{./S#1xS#2_bh_#3.png}
	\includegraphics[width=.48\textwidth]{./S#1xS#2_bh_#4.png}
	\caption{The ratio of radii $\beta$ as a function of the closing radius $r_0$ for both topologies with conformal boundary $S^#1\times S^#2$. The red diamond is the analytical calculation for the singular ratio $\beta_s$.}
	\label{fig:S#1S#2_bh_1}
	\end{figure}
}

\newcommand{\addFbFigureSingle}[4]{
	\begin{figure}[H]
	\centering
	\includegraphics[width=.8\textwidth]{S#1xS#2_Fb_#3.png}
	\FPeval{\result}{clip(round(((#2-1)/(#1-1))^(1/2),3))}
	\caption{The action $I$ as a function of the ratio of radii $\beta$ for both topologies with conformal boundary $S^#1\times S^#2$. The red diamond is the analytical calculation for the singular ratio $\beta_s=\result$, and the black line follows the dominant solution for each ratio $\beta$. We see that there is a first order phase transition at $\beta_c = #4$.}
	\label{fig:S#1S#2_Fb_1}
	\end{figure}
}

\newcommand{\addFbFigure}[5]{
	\begin{figure}[H]
	\centering
	\includegraphics[width=.48\textwidth]{./S#1xS#2_Fb_#3.png}
	\includegraphics[width=.48\textwidth]{./S#1xS#2_Fb_#4.png}
	\FPeval{\result}{clip(round(((#2-1)/(#1-1))^(1/2),3))}
	\caption{The action $I$ as a function of the ratio of radii $\beta$ for both topologies with conformal boundary $S^#1\times S^#2$. The red diamond is the analytical calculation for the singular ratio $\beta_s=\result$, and the black line follows the dominant solution for each ratio $\beta$. We see that there is a first order phase transition at $\beta_c = #5$.}
	\label{fig:S#1S#2_Fb_1}
	\end{figure}
}

\newcommand{\addgdbFigure}[4]{
	\begin{figure}[H]
	\centering
	\includegraphics[width=.48\textwidth]{./S#1xS#2_gd1b_#3.png}
	\includegraphics[width=.48\textwidth]{./S#1xS#2_gd1b_#4.png}
	\caption{$g_{(d)}^1$ as a function of the ratio of radii $\beta$ for both topologies with conformal boundary $S^#1\times S^#2$. This is proportional to the expectation value of the energy-momentum tensor. The red diamond is the analytical calculation for the singular ratio $\beta_s$.}
	\label{fig:S#1S#2_gd1b_1}
	\end{figure}
}

\begin{document}

\title{Generalized Hawking-Page transitions}
\date{\today}
\author{Ofer Aharony,}
\author{Erez Y. Urbach}
\author{and Maya Weiss}
\affiliation{Department of Particle Physics and Astrophysics, \\
	Weizmann Institute of Science, Rehovot 7610001, Israel}
\emailAdd{ofer.aharony@weizmann.ac.il}
\emailAdd{erez.urbach@weizmann.ac.il}

\abstract{
We construct holographic backgrounds that are dual by the AdS/CFT correspondence to Euclidean conformal field theories on products of spheres $S^{d_1}\times S^{d_2}$, for conformal field theories whose dual may be approximated by classical Einstein gravity (typically these are large $N$ strongly coupled theories). For $d_2=1$ these backgrounds correspond to thermal field theories on $S^{d_1}$, and Hawking and Page found that there are several possible bulk solutions, with two different topologies, that compete with each other, leading to a phase transition as the relative size of the spheres is modified. By numerically solving the Einstein equations we find similar results also for $d_2>1$, with bulk solutions in which either one or the other sphere shrinks to zero smoothly at a minimal value of the radial coordinate, and with a first order phase transition (for $d_1+d_2 < 9$) between solutions of two different topologies as the relative radius changes. For a critical ratio of the radii there is a (sub-dominant) singular solution where both spheres shrink, and we analytically analyze the behavior near this radius. For $d_1+d_2 < 9$ the number of solutions grows to infinity as the critical ratio is approached.
}

\maketitle

\section{Introduction and summary} \label{sec:introduction}

Gauge/gravity duality allows us to translate results in gravity to results in field theory, and vice versa. In particular, there is a large set of examples of $d$-dimensional conformal field theories (CFTs) whose dual is well-approximated by classical Einstein gravity on $AdS_{d+1}$ (times some compact manifold). These include the $4d$ ${\cal N}=4$ large $N$ $SU(N)$ supersymmetric Yang-Mills theory at strong `t Hooft coupling, and the large $N$ maximally supersymmetric $3d$ and $6d$ CFTs. For these theories classical computations in gravity can teach us about the behavior of the field theory (in the large $N$ limit). Moreover, since the gravitational sector is shared between many different field theories (which differ, for instance, in the choice of compact manifold), gravity computations give us universal results that are valid for many different field theories.

CFTs on some manifold ${\cal M}_d$ are dual by the AdS/CFT correspondence to gravitational theories on asymptotically anti-de Sitter (AdS) spaces, whose boundary is conformally equivalent to ${\cal M}_d$. The most studied examples are ${\cal M}_d=\mathbb{R}^d$ or ${\cal M}_d=S^d$, where the metric is exactly anti-de Sitter. Not many results are known about other spaces (see, for instance, \cite{Witten:1999xp,Anderson2004,deBoer:2004yu} for some general comments). One particularly interesting example is the Euclidean space ${\cal M}_d = S^{d-1}\times S^1$, first studied by Hawking and Page~\cite{Hawking1983}. This is interesting because it corresponds to the CFT on $S^{d-1}$ at finite temperature. Hawking and Page showed that there are two classes of smooth solutions with these boundary conditions. One set of solutions is given by anti-de Sitter space in global coordinates with Euclidean time identified; in this solution the $S^{d-1}$ factor shrinks to zero in the interior of space, such that the topology is $\mathbb{R}^d\times S^1$. The other set is described by Euclidean black holes, in which the $S^1$ shrinks to zero in the interior at the Euclidean horizon; this set has the topology $S^{d-1}\times \mathbb{R}^2$. The AdS/CFT correspondence identifies the partition function of the Euclidean field theory, in the gravity approximation, with the exponent of minus the Euclidean gravitational action (summed over all solutions with given boundary conditions). Hawking and Page showed that different classical solutions dominate the partition function at different temperatures (different ratios of the $S^1$ radius to the $S^{d-1}$ radius), with a phase transition between them at some critical temperature. When the dual field theory is a gauge theory, this phase transition may be identified (using various order parameters) with a confinement/deconfinement transition~\cite{Witten1998,Witten1998_2}.
On the field theory side, a sharp phase transition at finite volume is possible only when the number of degrees of freedom is infinite; for a finite number we expect a crossover. The translation of this to the gravity side is that quantum effects (corresponding to $\frac{1}{N}$ corrections in the field theory) will turn the phase transition into a crossover.

In this paper we generalize the discussion of Hawking and Page to CFTs on $\sd$. This no longer has a thermal interpretation, but the CFT partition function on such Euclidean manifolds is still an interesting observable. Similar observables are computed analytically using localization in supersymmetric CFTs~\cite{Pestun:2016zxk}. The result may also be related by analytic continuation to the physics of the canonical vacuum state in the CFT on the $d_1$-dimensional de Sitter space times $S^{d_2}$~\cite{Blackman:2011in}.
For this case we also find that there are two possible topologies, $\mathbb{R}^{d_1+1}\times S^{d_2}$ and $S^{d_1}\times \mathbb{R}^{d_2+1}$, and that there is a phase transition between them as the ratio of radii, $\beta \equiv R_{S^{d_2}}/R_{S^{d_1}}$ is varied. We were not able to find analytical solutions for general $\beta$, but we find numerical solutions for various values of $(d_1, d_2, \beta)$; the large isometry group of the space implies that the equations of motion are ordinary differential equations in a single variable. One important difference from the $d_2=1$ case is that here only one solution exists for $\beta \ll1$ and for $\beta \gg1$; there is no analog of the thermal AdS solution that exists for all ratios. In addition, there is in this case a solution with a conical singularity where both spheres shrink together, for some critical ratio $\beta_s$. This solution can be found analytically, and we can also analyze analytically the small fluctuations around it. For $d < 9$ we find oscillatory behavior of these fluctuations, such that the number of solutions with a given ratio $\beta$ grows as $\beta \to \beta_s$; however, the new solutions and the singular solution never dominate, and there is a first order phase transition between the solution which exists at $\beta \ll 1$ and the solution (of the other topology) which exists at $\beta \gg 1$, at some critical $\beta$ close to $\beta_s$. The case of $d \geq 9$ is unphysical, since there are no consistent quantum gravitational theories that have such $AdS_{d+1}$ solution with small curvature (indeed, even without requiring a gravity dual, there are so far no known interacting conformal field theories with $d \geq 9$). However, the classical analysis can be performed also in this case, and leads to a second order phase transition at $\beta=\beta_s$, with only one solution existing for every value of $\beta$.

\begin{figure}[t]
\centering
\includegraphics[width=.49\textwidth]{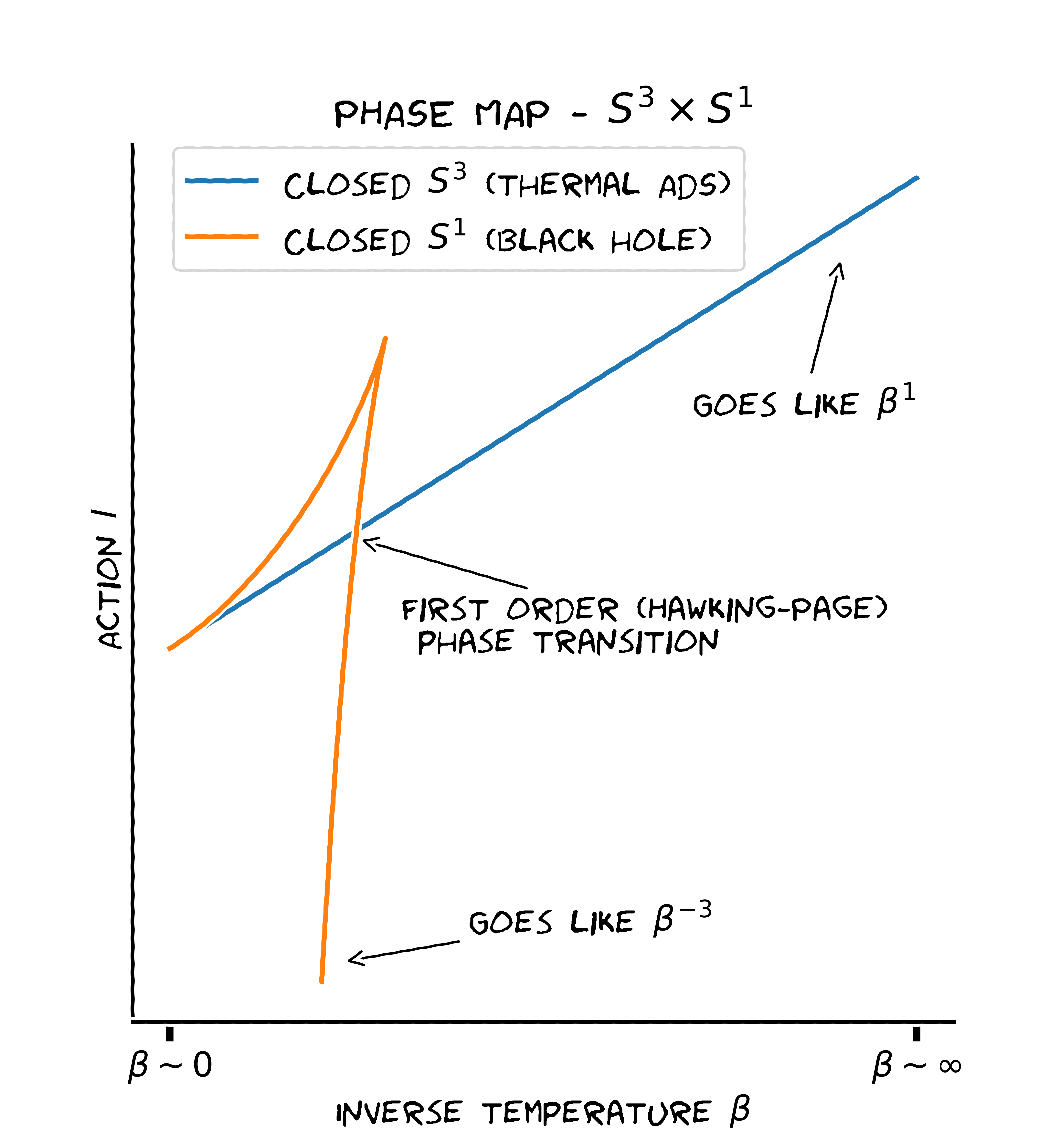}
\includegraphics[width=.49\textwidth]{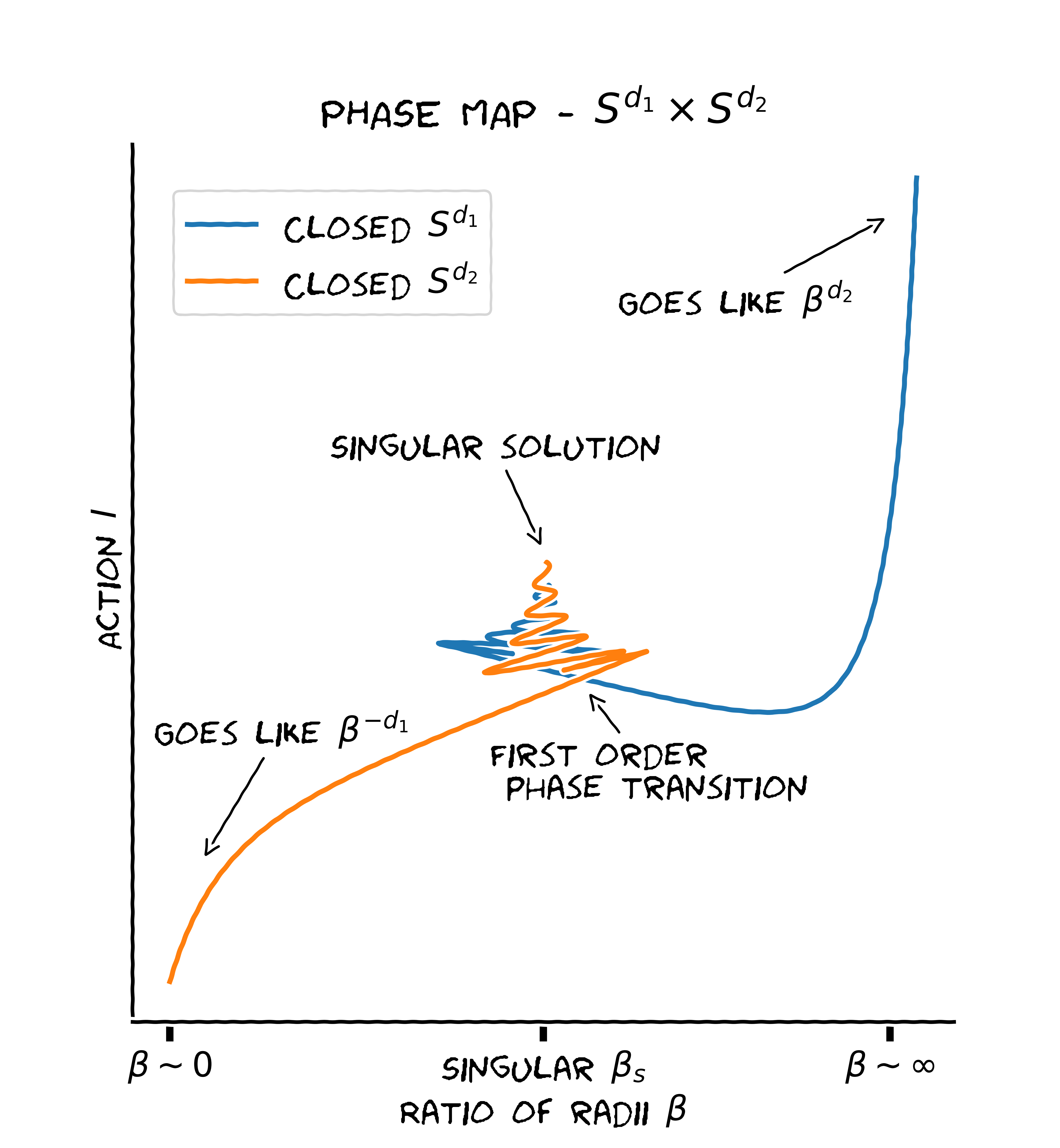}
\caption{Schematic plots of the Euclidean action $I$ as a function of the ratio $\beta$ between the two radii. On the left, for the Hawking-Page case $S^3\times S^1$, with $\beta=R_{S^1}/R_{S^3}$. On the right, for high-dimensional spheres $\sd $ ($d_1,d_2 \ge 2$, $d_1+d_2 < 9$), with $\beta=R_{S^{d_2}}/R_{S^{d_1}}$. Our analytic results do not fix the sign of $I$ in the asymptotic regions, we took the signs from the numerical results. As we note below (section \ref{sec:num_results} and section \ref{sec:num}) these signs are correct for $d_1 > d_2$. For $d_1=d_2$ the sign of $I$ in the two asymptotic regions is equal and is dimension-dependent.} 
\label{fig:schematic}
\end{figure}

The phase structure of the Hawking-Page case, and of the $\sd$ case with $d < 9$, are schematically summarized in figure \ref{fig:schematic}. The case of a CFT on $S^2\times S^d$ was already studied by the same methods in \cite{Blackman:2011in}, with the motivation of continuing the $S^d$ sphere to de Sitter space, and understanding thermal phase transitions in de Sitter space for confining theories arising from CFTs on $S^2$. The results we find for the phase structure exactly agree with theirs. As in the case of the Hawking-Page transition, quantum effects should turn the phase transition into a crossover, bu we do not discuss them here.

We begin in section \ref{sec:setup} by describing the setup and the observables we can compute. In section  \ref{sec:sing_sol_pert} we discuss in detail the singular solution and the perturbations around it. In section \ref{sec:hybrid_method_calc_energy} we present some details of our numerical implementation of the holographic renormalization which is needed for computing the Euclidean action. Our numerical results are presented in section  \ref{sec:num_results} for the specific case of $S^3\times S^2$, and then in section \ref{sec:num} we briefly present the results for some other representative examples.

The topology of space is different on the two sides of the phase transition, but in the pure gravity theory there is no simple order parameter that distinguishes them. If our bulk gravity theory contains $(d_2+1)$-dimensional extended objects which can end on the boundary, then these correspond to $d_2$-dimensional extended operators in the CFT; for example, this happens for $d_2=1$ when we have a string theory in the bulk, and strings ending on the boundary may be identified with Wilson loops. In such a case, the expectation value of the $d_2$-dimensional operator integrated on $S^{d_2}$ gives an order parameter for the phase transition, since one can find $(d_2+1)$-dimensional extended configurations in the bulk that end on the $S^{d_2}$, leading to a non-zero expectation value, if and only if the topology is of the form $S^{d_1}\times \mathbb{R}^{d_2+1}$. A similar discussion applies with $d_1 \leftrightarrow d_2$. In many examples of the AdS/CFT correspondence such order parameters exist, but their field theory interpretation depends on the specific case, and we will not discuss it here.


Our results give a universal behavior for all theories that are well-approximated by classical Einstein gravity in some limit. We consider solutions where only the gravitational field is turned on, and no other fields. Assuming that all other fields are set to zero at the boundary, it is natural to expect that they will vanish in the minimal-action solutions, but this may not be true in some examples, in which case our solutions would be sub-leading saddle points in these examples.

In many cases it is interesting to turn on additional bulk fields in addition to the metric. In particular, for supersymmetric theories, it is often interesting to compactify them on $\sd$ in a way that preserves supersymmetry, and this requires turning on additional fields. In this paper we discuss only solutions with no extra fields; in particular in supersymmetric theories, these solutions preserve no supersymmetry (since it is broken by the compactification of the field theory on $\sd$, in the absence of extra background fields). It would be interesting to look for solutions in which extra fields are turned on. There are also many other generalizations of our work, to different space-times, with products of more spheres, or with smaller isometry groups.

It would also be interesting to investigate other field theories, that do not have a dual described by classical Einstein gravity. These can include weakly coupled field theories, whose $\sd$ partition function can be computed in perturbation theory, and it would be interesting to compare their qualitative behavior to the one that we find here for strongly coupled field theories (for the $S^{d-1}\times S^1$ case, a similar transition exists at large $N$ also in weakly coupled gauge theories \cite{Sundborg,Hagedorn}).

\section{Setup and conventions} \label{sec:setup}

Our goal is to find Euclidean gravitational duals of CFTs on the space $\sd$, for $d$-dimensional CFTs ($d=d_1+d_2$) whose holographic dual is well-approximated by classical Einstein gravity on $AdS_{d+1}$ (perhaps multiplied by some compact space, which we assume nothing depends on so that it will play no role in our discussion). Namely, we look for $(d+1)$-dimensional manifolds that satisfy the (Euclidean) Einstein equations with a negative cosmological constant $\Lambda$, and have a conformal boundary $\sd$. We assume for simplicity that no additional fields are turned on except the metric; this gives us universal solutions that are present in all CFTs of this type, though they may not necessarily be the dominant solutions for specific CFTs. We will also assume the manifolds preserve the symmetries of the space that the CFT lives on\footnote{We expect the solutions that break the symmetry to have higher action and  not dominate the dynamics, though we have not proven this.}, and can thus can be foliated into $SO(d_1+1)\times SO(d_2+1)$ orbits. We can therefore write the metric as
\begin{equation}
	G_{\mu \nu} dx^{\mu} dx^{\nu}=dz^2+f^2(z)d\Omega_{d_1}^2+h^2(z)d\Omega_{d_2}^2. \label{eq:regular_coord}
\end{equation}
Here $d\Omega_{d_i}^2$ is the $SO(d_i+1)$-invariant metric of the unit sphere $S_{d_i}$, and $f,h$ may be thought of as the radii of each sphere for a given $z$. Here we chose a convenient reparameterization of the radial coordinate $z$, with the boundary at $z\rightarrow \infty$, and space ending smoothly at $z=0$ in the interior. 
The cosmological constant $\Lambda$ dictates the asymptotic behavior $f,h \propto e^\frac{z}{l}$ as 
$z\gg l \sim \Lambda^{-\frac{1}{2}}$, where $l$ is related to the AdS radius; we will work in units with $l=1$ from here on.
The range of $z$ is infinite; for our numerics it will be useful to work with a different coordinate $\rho=e^{-z}$ that will have a finite range in our solutions. In terms of this coordinate we write:
\begin{equation}
	G_{\mu \nu} dx^{\mu} dx^{\nu} = \frac{1}{\rho^2}\( d\rho^2 + r_1^2(\rho)d\Omega_{d_1}^2+ r_2^2(\rho)d\Omega_{d_2}^2 \) . \label{eq:compact_metric}
\end{equation}
This is also the convention used in the Fefferman-Graham expansion in order to renormalize the gravitational action~\cite{DeHaro2000}. This time $\rho=0$ is the boundary, and we expect the functions $r_1,r_2$ to stay finite along the entire compact region. We are going to use both conventions in this paper. At the beginning we use mostly the $f,h$ convention. Starting from the end of section \ref{sec:sing_sol_pert} we switch to the $r_1,r_2$ convention to calculate asymptotic observables.

As in the Hawking-Page transition, we expect the solutions where space ends smoothly at $z=0$ to associate with two types of topologies -- $\mathbb{R}^{d_1+1}\times S^{d_2}$ and $S^{d_1}\times\mathbb{R}^{d_2+1}$. Because the entire discussion is symmetric in $(d_1,d_2)$, it is enough to conisder the case of $\mathbb{R}^{d_1+1}\times S^{d_2}$. In this case the $S^{d_1}$ sphere shrinks to zero size at $z=0$ ($\rho=1$), $f(z=0)=r_1(\rho=1)=0$.

Aside from the topology, the solution should also stay regular at the closing point $z=0$. 
This determines $f^2(z) = z^2 + O(z^4)$ and $h^2(z)=r_0^2+O(z^2)$ for some ``closing radius" $r_0>0$. These conditions can be written as boundary conditions for the fields $f,h$ at $z=0$:
\begin{align}
	f(z=0)=0, \quad f'(z=0)=1, \nonumber \\
	h(z=0)=r_0, \quad h'(z=0)=0. \label{eq:initial_cond}
\end{align}
One way to derive these conditions is to look at the expression for the Kretschmann scalar $K=R_{\mu\nu\rho\sigma}R^{\mu\nu\rho\sigma}$. For the metric \eqref{eq:regular_coord} it is
\begin{align}
	K  &= d_1\left(\frac{f''}{f}\right)^{2}+d_1\left(d_1-1\right)\left(\frac{1-\left(f'\right)^{2}}{f^{2}}\right)^{2}+d_1d_2\left(\frac{f'h'}{fh}\right)^{2}\nonumber\\
	& \qquad + d_2\left(\frac{h''}{h}\right)^{2}+d_2\left(d_2-1\right)\left(\frac{1-\left(h'\right)^{2}}{h^{2}}\right)^{2}.
\end{align}
This expression is a sum of squares. Requiring each square to stay finite (for $d_1>1$) separately at $z=0$ gives the boundary conditions \eqref{eq:initial_cond}. Note that for small closing radius $r_0$ the Kretschmann scalar behaves as $K(z=0) \sim r_0^{-2}$, so as $r_0 \to 0$ the curvature blows up, as expected since for $r_0=0$ we have a conical singularity at $z=0$ where both spheres shrink together. In summary, the most general boundary condition is parametrized by a choice of topology (which sphere closes at $z=0$) and of the closing radius $r_0>0$ (of the other sphere at $z=0$).

The Einstein equations for a $D$-dimensional space in vacuum with a cosmological constant $\Lambda$ are 
\begin{equation}
	R_{\mu \nu} = \frac{2\Lambda}{D-2} G_{\mu \nu}, \label{eq:einstein_eqs}
\end{equation}
where in our case $D=d_1+d_2+1$.
In our choice of units we have $\frac{2\Lambda}{D-2}=-(D-1)=-d$. Substitution of the metric $\eqref{eq:regular_coord}$ gives the following second order differential equations for $f,h$:
\begin{align}
\frac{f''}{f}+d_2\frac{f'}{f}\frac{h'}{h}+\left(d_1-1\right)\frac{\left(f'\right)^{2}-1}{f^{2}}	=	d, \nonumber \\
\frac{h''}{h}+d_1\frac{f'}{f}\frac{h'}{h}+\left(d_2-1\right)\frac{\left(h'\right)^{2}-1}{h^{2}}	=	d, \label{eq:eqs_of_motion}
\end{align}
and also a constraint coming from the equation for $G_{zz}$ (which we fixed by a choice of diffeomorphism):
\begin{equation}
	d_1 \frac{f''}{f}+d_2 \frac{h''}{h} = d. \label{eq:restriction}
\end{equation}
This equation can be derived from the first two \eqref{eq:eqs_of_motion} assuming it is satisfied at one point. Moreover, the constraints of regularity at the closing point \eqref{eq:initial_cond} together with the equations of motion \eqref{eq:eqs_of_motion} lead to the constraint being satisfied at the closing point. Thus it is enough to solve \eqref{eq:eqs_of_motion} with these initial conditions, and the constraint will be satisfied.

For a given topology and $r_0$ there is a unique metric given by solving \eqref{eq:eqs_of_motion} with the initial conditions \eqref{eq:initial_cond}, giving the pair of functions $f(z),h(z)$. This metric is a saddle point of the AdS partition function, and so can be used to extract CFT information. Generally, it is convenient to write metrics on asymptoticaly AdS spaces as
\begin{equation}
	G_{\mu \nu} dx^{\mu} dx^{\nu} = \frac{1}{\rho^{2}}\( d\rho^{2} + g_{ij}(x,\rho) dx^{i}dx^{j}\) ; \label{eq:norm_metric}
\end{equation}
in our case this is exactly the metric written in terms of $r_1(\rho),r_2(\rho)$ \eqref{eq:compact_metric}. We then have the Fefferman-Graham expansion of $g(x,\rho)$ around the boundary $\rho=0$ ~\cite{DeHaro2000}, which in even dimensions takes the form
\begin{equation}
	g(x,\rho) = g_{(0)}(x)+\rho^2 g_{(2)}(x) +\cdots +\rho^d 
	\( g_{(d)}(x) + h_{(d)}(x) \log (\rho) \) + \cdots,\label{eq:metric_exp}
\end{equation}
and in odd dimensions
\begin{equation}
	g(x,\rho) = g_{(0)}(x)+\rho^2 g_{(2)}(x) + \cdots +\rho^{d-1} g_{(d-1)}(x)+ \rho^d g_{(d)}(x) + \cdots. \label{eq:metric_exp2}
\end{equation}

In the AdS/CFT correspondence, such solutions describe CFTs on a space conformally equivalent to $\go$. In our case the CFT metric is defined up to conformal transformations by the ratio between the two spheres' radii. In terms of the bulk metric it is
\begin{equation}
	\beta = \lim_{z\rightarrow\infty} \frac{h(z)}{f(z)}. \label{eq:beta_def}
\end{equation}
Note that in the case of $d_2=1$ (Hawking-Page), $\beta$ has the interpretation of the inverse temperature of the thermal CFT on the unit sphere $S^{d_1}$. Given a solution with some $r_0$ we can compute $\beta$, which tells us which theory our saddle-point metric $G$ corresponds to. 

In order to find the phase diagram of the theory we need to find all the saddle-point metrics with the same asymptotic geometry $\go$. We can label the solution by the triad $(d_1,d_2,\beta)$. This label is not unique since several solutions (corresponding to different $r_0$'s) might have the same asymptotic $\beta$. Moreover, in this notation a solution with a triad $(d_2,d_1,\frac{1}{\beta})$ also has the same asymptotic geometry, and all those solutions are saddle points of the same path integral. Therefore properties of the theory should be compared between all solutions with triads of both kinds (if they exist).

The $d$'th order term of the expansion \eqref{eq:metric_exp}, $\gd$, gives (up to a constant proportionality number) the expectation value of the CFT stress tensor $g^{(d)}_{ij} \sim \left<T_{ij}\right>$ (for even dimensions there is a further anomalous contribution to the trace of $\langle T_{ij} \rangle$). In our case this means the stress tensor expectation values are given by $\frac{d^d}{d\rho^d} (r_{1,2}^2(\rho))|_{\rho=0}$.

In the AdS/CFT correspondence, the partition function of the CFT may be approximated as $Z \simeq \exp(-\frac{1}{16\pi G_N}I)$, where $I$ is 
the Euclidean action of the saddle point metric\footnote{Here we took the coefficient $\frac{1}{16\pi G_N}$ outside the action; in our $l=1$ units this is a very large dimensionless number. In our computations it only appears multiplying $I$ everywhere, and we will leave it implicit.}. In order to calculate it, one needs to regularize the naive Hilbert-Einstein action by introducing a boundary at some cutoff $\rho=\epsilon$ and subtracting covariant boundary counter-terms,
\begin{equation}
	I[G] = \int_M d^D  x \sqrt{G} R -\int_{\partial M}d^{d}x\sqrt{\gamma}2K - \int_{\partial M} d^{d}x \sqrt{\gamma} {\cal L} _{ct}[\gamma], \label{eq:holo_reg_sketch}
\end{equation}
where $\gamma$ is the metric induced on the boundary~\cite{DeHaro2000}. 
In general we need to sum the contributions to $Z$ from different saddle points; in the large $N$ limit the actions are very large so we can ignore any subleading saddle points.
Given an asymptotic geometry, the free energy (defined as minus the log of the partition function) is the minimum between the actions of all the different configurations with either $(d_1,d_2,\beta)$ or $(d_2,d_1,\frac{1}{\beta})$. We denote this function by $I(\beta)$, and it encodes the different phases of the CFT. In the case of $d_2=1$, after dividing by $\beta$, this quantity is just the thermal free energy as a function of the inverse temperature.

\section{The singular solution and its perturbations} \label{sec:sing_sol_pert}

Some qualitative properties of the phase diagram are apparent without calculation. In the limit $\beta\rightarrow 0$, where $r_2\ll r_1$ at the boundary, we expect the smaller sphere to shrink much before the larger one, such that the only solution would have the topology $S^{d_1}\times\mathbb{R}^{d_2+1}$ with a large $r_0$.\footnote{This statement not true when $d_2 = 1$. In that case there is a solution with closed $S^{d_1}$ sphere even as $\beta\rightarrow 0$, which is thermal AdS (see figure \ref{fig:schematic}), because this happens to be an exact solution for all $\beta$.} We also expect the CFT on such a space to behave like some $d_1$-dimensional theory on $S^{d_1}$, which is obtained by KK reduction on the other sphere. Thus, to first order the action should scale like the normalized volume of $S^{d_1}$, which is $\beta^{-d_1}$. Symmetrically, for $\beta\rightarrow \infty$ the solution will be topologically $\mathbb{R}^{d_1+1}\times S^{d_2}$ with a large $r_0$, and the action will scale like $\beta^{d_2}$. So for both $\beta \sim 0,\infty$ the action diverges, each limit as a result of a different bulk topology. Consequently, at least one phase transition is expected between the two topologies. Naively one may expect that decreasing $r_0$ decreases the asymptotic radius of the corresponding sphere, until some critical radius where $r_0=0$, where this branch of solutions connects to the branch with the other topology. In this case there would be just a single solution for every radius in the CFT, and a smooth transition between the two branches at $r_0=0$. We will see that this is true when $d_1+d_2 \geq 9$, but not in the physically interesting range of dimensions $d_1+d_2 < 9$, where the asymptotic radius will not be monotonic in $r_0$. In this range more than one solution of the same topology can exist for the same asymptotic radii. Recall that this happens already for $d_2=1$, where both ``large'' and ``small'' AdS black holes can exist with the same temperature.

Lacking a complete analytical solution, we turn to study the behavior for $r_0=0$.
By the analysis above (section \ref{sec:setup}) $r_0=0$ corresponds to a solution with a singular curvature at the closing point $z=0$ \footnote{In this paper we assume that the higher curvature corrections to Einstein's equations are small, and we neglect them. This will not be true when both $r_0$ and $z$ are very small, so our solutions cannot be trusted there. The features of the solutions that we describe will be independent of the form of these corrections for $d<9$. For $d \geq 9$, where the singular solution is dominant for some value of $\beta$, the corrections will be important in order to determine the precise nature of the phase transition; however, this case is unphysical.}. 
To find the solution we take the ansatz $h(z) = \alpha f(z)$. Substitution in \eqref{eq:eqs_of_motion},\eqref{eq:restriction} gives (for $d_1,d_2 > 1$) the solution
\begin{align}
	f_s(z) = \sqrt{\frac{d_1-1}{d-1}}\sinh(z), \quad h_s(z) = \sqrt{\frac{d_2-1}{d-1}}\sinh(z), \label{eq:singular_metric}
\end{align}
or in terms of $r_1,r_2$
\begin{align}
	r_1^s(\rho) = \sqrt{\frac{d_1-1}{d-1}} \frac{1-\rho^2}{2}, \quad
	r_2^s(\rho) = \sqrt{\frac{d_2-1}{d-1}} \frac{1-\rho^2}{2}.
	\label{eq:singular_metric_compact}
\end{align}
This solution has the ratio of radii
\begin{equation}
	\beta_s = \sqrt{\frac{d_2-1}{d_1-1}}. \label{eq:singular_temp}
\end{equation}
Since $d_1,d_2 > 1$, we have $d \geq 4$. In the singular solution $r_1^2$ and $r_2^2$ are fourth order polynomials in $\rho$, so for $d>4$
the $d$'th derivative of the metric at the bondary vanishes,
\begin{equation}
	g^s_{(d)} = 0. \label{eq:no_gd_singular}
\end{equation}
\eqref{eq:singular_metric_compact} is an explicit solution for $\beta=\beta_s$, but there may also be others so it may not be the dominant one.

Solutions with non-zero $r_0$ will look very different, and be non-singular, near the origin. But we expect that for small $r_0 \ll 1$, and far from the origin $z \gg r_0$, they will be small perturbations of the singular metric.
Thus, following \cite{Kol2002,Kalisch2017}, we look for small perturbations of \eqref{eq:singular_metric}, keeping only the terms linear in the perturbation in the equations of motion. 
Given a metric perturbation $\delta G^s$, we can look at its trace with respect to the unperturbed singular metric ${\rm tr}(G_s^{-1} \delta G^s)=G_s^{\mu \nu} \delta G^s_{\mu \nu}$.
The only trace-full perturbation turns out to be a coordinate change (a shift of the closing point). The other perturbation is traceless:
\begin{equation}
	\delta f_s(z) = \frac{1}{d_1} f_s(z) a(z), \qquad \delta h_s(z) = -\frac{1}{d_2} h_s(z)  a(z).
\end{equation}
Linearizing the Einstein equations \eqref{eq:eqs_of_motion} in $a(z)$ around the singular solution gives
\begin{equation}
	a^{\prime\prime}(z) + d \coth (z) a^\prime(z) + 2(d-1) \sinh^{-2}(z)a(z)=0. \label{eq:traceless_de}
\end{equation}
The two solutions of \eqref{eq:traceless_de} are 
\begin{equation}
	a_\pm(z) = \ _2 F_1 \( \frac{1}{2} {\alpha_\pm}, \frac{1}{2} \( 1+{\alpha_\pm} \) ; {\alpha_\pm} + \frac{1}{2} \( d+1 \) ; \tanh^2 (z) \) \tanh^{\alpha_\pm} (z), \label{eq:pert_sols_general}
\end{equation}
where $\alpha_\pm = -\frac{1}{2} \( d -1\pm \sqrt{(d-1)(d-9)} \) $. The general traceless perturbation is a linear combination of the two. We are mostly interested in the case of $d<9$, where $\alpha_\pm$ are complex and the two solutions are complex conjugates. In this case we can parameterize the perturbation using a complex number $a_0\in \mathbb{C}$ by 
\begin{equation}
	\delta f_s(z) = \frac{1}{d_1} f_s(z) \( a_0 \cdot a(z) + c.c. \) \qquad \delta h_s(z) = -\frac{1}{d_2} h_s(z) \( a_0 \cdot a(z) + c.c. \) , \label{eq:pert}
\end{equation}
where we take 
\begin{equation}
	a(z) = \ _2 F_1 \( \frac{1}{2} \alpha, \frac{1}{2} \( 1+\alpha \) ; \alpha + \frac{1}{2} \( d+1 \) ; \tanh^2 (z) \) \tanh^\alpha (z), \label{eq:pert_sols}
\end{equation}
and $\alpha = \alpha_+ = -\frac{1}{2} \( d -1+ \sqrt{(d-1)(d-9)} \) $.

For each such solution we can find the change in the asymptotic ratio of radii \eqref{eq:beta_def}; using \eqref{eq:pert} we find at first order in $a_0$:
\begin{equation}
	\frac{\delta \beta}{\beta_s} = - \( \frac{1}{d_1} + \frac{1}{d_2}\) \( a_0 \cdot a(\infty) +c.c. \) , \label{eq:pert_temp}
\end{equation}
where $a(\infty) = \ _2 F_1 \( \frac{1}{2} \alpha, \frac{1}{2} \( 1+\alpha \) ; \alpha + \frac{1}{2} \( d+1 \) , 1 \) $.

We can also find the perturbed $\gd$, by deriving the metric $g(\rho)$ $d$ times with respect to the coordinate $\rho=e^{-z}$. 
To first order in $a_0$
\begin{align}
	\delta g(\rho) = \frac{\( 1-\rho^2 \) ^2}{2(d-1)}  \( \frac{d_1-1}{d_1 } d\Omega^2_{d_1}  - \frac{d_2-1}{d_2} d\Omega^2_{d_2} \) (a_0 \cdot a(\rho) +c.c.).
\end{align}
Note that to first order the volume at $\rho=0$ is not modified because we perturbed by a traceless perturbation. The $d$'th derivative is then
\begin{align}
	\delta \gd = & a_0\cdot \( a^{(d)}(0) -4\binom{d}{2} a^{(d-2)}(0) + 24 \binom{d}{4} a^{(d-4)}(0) \) \\
	& \cdot \frac{1}{2(d-1)} \( \frac{d_1-1}{d_1 } d\Omega^2_{d_1}  - \frac{d_2-1}{d_2} d\Omega^2_{d_2} \) +c.c. \label{eq:pert_gd}
\end{align}
where $a^{(n)}(0)$ is the $n$-th derivative of $a(\rho) = a(z=-\log (\rho))$ at $\rho=0$. 

To calculate the perturbation of the action, we use the fact that the action is (an integral of) a scalar functional of $g(\rho)$ and of its derivatives at $\rho=0$. The singular solution $g_s(\rho)$ is a constant metric multiplied by a scalar function of $\rho$. Therefore the variation of the action at the singular solution is proportional to $g_{(0)s}^{-1}$. It follows that the variation with respect to the traceless perturbation is zero at leading order in $a_0$, because
\begin{equation}
 	\delta I = tr \( \frac{\delta I}{\delta \go}\Bigg|_{g_s} \delta g_{(0)s} \)
 	\propto tr \( g_{(0)s}^{-1} \delta g_{(0)s}\) = 0. \label{eq:pert_F}
 \end{equation} 

Our numerical solutions depend on the initial condition $r_0$, so in order to relate them to our discussion
we need to say something about the relation between $a_0$ and $r_0$. In our conventions the AdS radius is of order $1$. 
When $r_0\ll1$ we expect the solution around $z\sim r_0$ to be the same as a flat space solution, namely a solution where we ignore the cosmological constant term and expand in $r_0$ around the conical singularity (at $r_0=0$) in flat space. For such a solution $r_0$ is the only length scale.
Define $a_0(r_0)$ to be the value of $a_0$ such that the dimensionless perturbation to the singular solution, $a_0(r_0) \cdot a(z)$ \eqref{eq:pert}, behaves similar to the $r_0$ solution for $z > r_0$. Dimensional analysis implies that $a_0(r_0) \cdot a(z)$ has to be a function of the dimensionless ratio $\frac{z}{r_0}$ alone. By \eqref{eq:pert_sols}, for  $r_0 \ll z \ll 1 $ we get $a(z) \simeq z^{\alpha}$, and thus $a_0(r_0) = C r_0^{-\alpha} + O(r_0^{1-\alpha})$. This conclusion satisfies the consistency check $-{\rm Re}( \alpha) = \frac{d-1}{2} >0$, so that one gets back the singular solution in the limit $r_0 \rightarrow 0$. For $d<9$ the exponent is complex ${\rm Im}( \alpha) = \sqrt{(d-1)(9-d)}$, and $a_0(r_0)$, and thus also $\delta \beta(r_0)$ \eqref{eq:pert_temp}, has decayed oscillations to zero,
\begin{equation}
\delta \beta(r_0) \simeq {\tilde C} r_0^{\frac{d-1}{2}} \cos(\sqrt{(d-1)(9-d)} \log(r_0) + \phi),
\end{equation}
for some constants ${\tilde C}$ and $\phi$. We do not know how to compute these dimensionless constants, but we can extract them from our numerical solutions, discussed below.

From the analysis above we can draw a schematic picture of the phase map (figure \ref{fig:schematic}). Around $\beta \sim 0,\infty$ the action diverges like $\beta^{-d_1}, \beta^{d_2}$ respectively (we take $\beta=\frac{r_2(0)}{r_1(0)}$), each divergence comes from a different bulk topology. The two branches coincide at $\beta_s$ and at the same action. From \eqref{eq:pert_temp} we expect decayed oscillations of $\beta$ toward $\beta_s$ as $r_0\rightarrow 0$. Because $\delta I =0$ \eqref{eq:pert_F}, the two topologies approach $\beta_s$ in $I(\beta)$ with zero slope. The minimal action curve is expected to have at least one first order phase transition, close to $\beta_s$ on the $\beta$ axis (see figure \ref{fig:schematic}). Note that the second order variation of the action with respect to the deviation from the singular solution dictates whether there will be an infinite amount of phase transitions, which would happen if the action decreases as the solution becomes singular. In figure \ref{fig:schematic} we assumed that the action increases, so that we get a single first order phase transition; this scenario will be confirmed by our numerical analysis. Finally, by \eqref{eq:pert_gd} $\gd$ oscillates in a similar fashion to the ratio of radii but with (presumably) different phase and amplitude. Therefore the function $\gd(\beta)$ is expected to have spiral behavior around the point $(\beta_s,0)$.

In the unphysical case of $d\geq 9$ $\beta(r_0)$ behaves monotonically (at least around $\beta_s$, and for $\beta\sim 0,\infty$), so we expect to have
a second order phase transition that smoothly interpolates between the two topologies.

\section{Holographic numerical calculations} \label{sec:hybrid_method_calc_energy}

In order to find solutions for general $r_0$ (general $\beta$) we take initial conditions with many different values of $r_0$, and numerically solve the differential equations \eqref{eq:eqs_of_motion}.

Given a numerical solution $G$ for the equations of motion \eqref{eq:eqs_of_motion} we would like to calculate its Euclidean action. As mentioned above \eqref{eq:holo_reg_sketch}, the Hilbert-Einstein action should be regularized and renormalized by local boundary counter terms in order to be finite. The expression for the action is
\begin{equation}
	I[G] = \lim_{\varepsilon \rightarrow 0} \left[ \int_\varepsilon^1 d\rho \ L_{reg}(\rho) + I_{ct}(\varepsilon) \right] \label{eq:exact_action},
\end{equation}
where (for $d\le 6$)~\cite{DeHaro2000}
\begin{align}
	L_{reg}(\rho) &= \int d^d x \sqrt{G} \( R[G] +2\Lambda \) \label{eq:L_reg_def} \\
	I_{ct}(\varepsilon) & = \int_{M_\varepsilon} \sqrt{\gamma}\left[ 2(1-d) + \frac{1}{d-2} R + \frac{1}{(d-2)^2 (d-4)} \( R^{ij} R_{ij} + \frac{d}{4(d-1)}R^2 \)  \right] \label{eq:I_ct_def}.
\end{align}
In \eqref{eq:L_reg_def} the metric is at the point $(x,\rho)$, whereas in \eqref{eq:I_ct_def} $\gamma$ (and the curvatures in the integrand) is the induced metric from $G$ on the sub-manifold $M_\varepsilon$ at $\rho=\varepsilon$. Each term in \eqref{eq:I_ct_def} appears only for high enough dimension. The first appears for any $d$, the second for $d>2$ and the third for $d>4$. The counter-terms remove the diverging part of the regularized action, so that the limit of \eqref{eq:exact_action} gives a finite answer.

Unfortunately, a naive numerical calculation of \eqref{eq:exact_action} leads to a significant numerical error arising from the subtraction of two large and close numbers. One possible solution is to write the sum using one integrand. Defining $I_{reg}(\varepsilon) = \int_\varepsilon^1 d\rho \ L_{reg}(\rho)$, we obtain
\begin{equation}
	I_{reg} + I_{ct} = \int_\varepsilon^1 d\rho \ \( L_{reg}(\rho) - \partial_\rho I_{ct}(\rho) \) + I_{ct}(\varepsilon=1).
\end{equation}
The new integrand opens up a different problem: $\partial_\rho I_{ct}$ might diverge (depending on $d_1,d_2$) around the closing point $\rho\sim 1$. For this reason we use the fact that both integrands are finite in the bulk, and write instead
\begin{equation}
	I_\text{numerical}^{[\varepsilon,1]} = \int_\varepsilon^\delta d\rho \ \( L_{reg}(\rho) - \partial_\rho I_{ct}(\rho) \) + \int_\delta^1 d\rho \ L_{reg}(\rho) + I_{ct}\Big|_{\rho=\delta}, \label{eq:try_delta}
\end{equation}
where $\delta$ was chosen such that the first and last terms are comparably small.
Although this solution reduces the error around the boundary, it is not enough to eliminate it to a sufficient degree.
Close to the boundary the two integrands (of the first integral in \eqref{eq:try_delta}) diverge differently, and so the complete integral still carries a large numerical error in the limit $\varepsilon \rightarrow 0$. 
Nonetheless, we can trust the value of the numerical integral \eqref{eq:try_delta} for $\varepsilon$ which is not too small.

To find the value of $I^{[0,1]}$ with better accuracy we can incorporate the 
Fefferman-Graham expansion of $g(\rho)$ around $\rho=0$ \eqref{eq:metric_exp},\eqref{eq:metric_exp2}. We can expand the integrand of the renormalized action $L_{reg}-\partial_\rho I_{ct}$ up to some positive power $\rho^n$ of $\rho$ as a local functional of $g_{(0)},g_{(d)}$ (the entire expansion of the function $g(\rho)$ in \eqref{eq:norm_metric} may be written using local functionals of $\go,\gd$). Taking the integral between $0$ and $\varepsilon$ will give the correct $I^{[0,\varepsilon]}$ up to the positive power $\varepsilon^{n}$. We can denote it $I^{[0,\varepsilon]}_\text{analytical} = I^{[0,\varepsilon]} + O(\varepsilon^n)$. This expression is still a local function of $g_{(0)},g_{(d)}$. Their value can be found numerically by fitting the numerical solution in some range between $0$ and some small $\rho=\varepsilon_\text{sampling}$ to the analytical expansion of the metric, up to the positive order $\rho^{d+n}$.
The combined calculation of the action can be written as
\begin{equation}
	I_{ren}^{[0,1]} = I^{[0,\varepsilon]}_\text{analytical}[g_{(0)}^{num},g_{(d)}^{num}] + I_\text{numerical}^{[\varepsilon,1]}. \label{eq:energy_calc}
\end{equation}
We note that there is a trade-off in choosing $\varepsilon$. Making it too small 
will bring back the numerical error caused by the integrands' divergence in $I_\text{numerical}^{[\varepsilon,1]}$ \eqref{eq:try_delta}. Making it too large will result in a larger error in $I^{[0,\varepsilon]}_\text{analytical}$, as it was calculated with error $O(\varepsilon^n)$.

In the convention we chose for the coordinates \eqref{eq:norm_metric} the solution's closing point is at $\rho=1$. Choosing a different closing point is equivalent to taking a different boundary metric $\go$, related by a conformal transformation. For even dimensions, due to the conformal anomaly, this choice affects the action, so we need to fix it when comparing different solutions. We will choose our CFT coordinates such that the total CFT volume $\int d^d x \sqrt{\go} $ is constant, independent of $\beta$. Our numerical solutions generally give a different volume, and to fix this we use the transformation $I[\exp(2\sigma)\go]=I[\go]+\int d^d x {\cal A}(x) \sigma (x)$. The scale anomaly ${\cal A}(x)$, which is non zero only for even dimensions, is known analytically as a local functional of the boundary metric $\go$~\cite{Henningson1998}. 
We thus need to add an extra factor
\begin{equation}
	I[g] = I_{ren}^{[0,1]}[g] + 2\int d^d x {\cal A}(\go^{num}(x)) \log\( \sqrt{\go^{num}}\) ,
\end{equation}\newline
and this gives our final result which we plot below. Note that the additional terms ensures that we still have $\delta I = 0$ \eqref{eq:pert_F} at leading order in the deviation away from the singular solution.

Several arbitrary parameters were introduced above to carry out the calculation of the action. Here is a list with the explanation of each of them, together with the measured effect of altering their value.
\begin{itemize}
	\item $\delta$ - used to specify the range of integration for the counter-terms \eqref{eq:try_delta}.
	Varying $\delta$ has almost no effect on the final result.
	\item $\varepsilon_\text{sampling}$ - the range $[0,\varepsilon_\text{sampling}]$ from which the fit to the analytical expansion, determining $\go$ and $\gd$, was taken. Varying $\varepsilon_\text{sampling}$ has almost no effect on the final result.
	\item $\varepsilon$ - circumscribes between the analytical and the numerical integration. $\varepsilon$ needed some fine-tuning, to minimize the numerical errors.
\end{itemize}
The figures we draw below contain estimates of the numerical errors, though in most cases these are too small to be visible. These can be reduced by increasing the accuracy of the numerical analysis.

\section{Summary of numerical results for \texorpdfstring{$S^3 \times S^2$}{}} \label{sec:num_results}

As we described in section \ref{sec:setup}, for a given CFT metric $\go$ we need to consider all the bulk solutions with that boundary metric. 
We now label the CFT geometry by the ratio $\beta$ of the $S^{d_2}$ radius to the $S^{d_1}$ radius, where $d_1>d_2$. Note that in the notations of section \ref{sec:setup} we need to consider solutions both with a closed $S^{d_1}$ sphere and ratio $\beta$, and with a closed $S^{d_2}$ sphere and ratio $\frac{1}{\beta}$.
In the case of $d_1=d_2$ we can consider the same solution under both $\beta$ and $\frac{1}{\beta}$. In this section we will describe the features found in the numerical analysis for various $d_1,d_2$ through the example of $S^3\times S^2$. Some plots for other dimensions are given in the next section.

\subsection{The ratio of radii as a function of the closing radius}

The analytical examination above (section \ref{sec:sing_sol_pert}) suggests that for large $r_0\gg1$ the open sphere in the IR would stay bigger along the bulk, and the final ratio of radii $\beta$ would be very large or very small (depending on the convention for the ratio). For small $r_0 \ll 1$ a complex critical exponent was predicted near $\beta_s=\sqrt{\frac{d_2-1}{d_1-1}}$ \eqref{eq:singular_temp}. In figure \ref{fig:S3S2_bh_1} we can see exactly those characteristics. Similar behavior can be found also for other $d_1,d_2$ (see section \ref{sec:num}). The analytical description captures all the qualitative features appearing in the figures.

The behavior of $\beta(r_0)$ around the singular solution was found above to take the form
$\beta(r_0)-\beta_s \propto r_0^\rho \cos(\omega \log(r_0) + \phi)$ where $\rho = \tfrac{1}{2}\( d -1 \) $ and  $\omega = \sqrt{(d-1)(9-d)}$. In order to check this result we fitted the log of the calculated $(\beta(r_0)-\beta_s)$ to the aforementioned equation, to find $\rho,\omega$. The fit was taken with the function \texttt{FindFit} of \texttt{Mathematica} with 100 data points, for $r_0$ between $e^{-1}$ and $e^{-7}$ (which is also the lower bound for accurate results of $\beta$ using the Mathematica integrator). The results are presented in table \ref{tb:temp_exponent}. The numerical exponent agrees with the analytical one to high precision. It is evident that the error in the fit grows with the total dimension $d$. This is because the frequency becomes smaller as $d$ gets closer to $9$, making it more difficult to estimate it over the same data range. As expected, no oscillations were found for $d \ge 9$ (not shown in the table).
\begin{table}[ht]
	\begin{center}
	    \begin{tabular}{|c|c|c|c|c|c|}
	    \cline{2-5}
	    \multicolumn{1}{c}{} & \multicolumn{2}{|c|}{Expected} & \multicolumn{2}{c|}{Fitted}\\ \cline{1-5}
	    $d_1,d_2$ & $\rho$ & $\omega$& $\rho$ & $\omega$\\  \cline{1-5}
	    $2,3$ & $2.$ & $2.$ & $1.93715$ & $1.9773$ \\
	    $2,4$ & $2.5$ & $1.93649$ & $2.40413$ & $1.94027$ \\
	    $2,5$ & $3.$ & $1.73205$ & $2.88297$ & $1.65561$ \\
	    $2,6$ & $3.5$ & $1.32288$ & $3.23972$ & $1.2337$ \\
	    $3,2$ & $2.$ & $2.$ & $2.02449$ & $1.99631$ \\
	    $3,3$ & $2.5$ & $1.93649$ & $2.51484$ & $1.96453$ \\
	    $3,4$ & $3.$ & $1.73205$ & $2.98307$ & $1.73993$ \\
	    $3,5$ & $3.5$ & $1.32288$ & $3.38868$ & $1.2696$ \\
	    $4,2$ & $2.5$ & $1.93649$ & $2.60947$ & $1.92624$ \\
	    $4,3$ & $3.$ & $1.73205$ & $3.08959$ & $1.68927$ \\
	    $4,4$ & $3.5$ & $1.32288$ & $3.47092$ & $1.03088$ \\
	    $5,2$ & $3.$ & $1.73205$ & $3.0191$ & $1.70681$ \\
	    $5,3$ & $3.5$ & $1.32288$ & $3.31987$ & $0.909707$ \\
	    $6,2$ & $3.5$ & $1.32288$ & $3.27731$ & $0.484668$ \\
	    \cline{1-5}
	   
	    \end{tabular}
	\end{center}
	\caption{Fitted values for the complex exponent of $\beta(r_0)$ vs the analytical value.}
	\label{tb:temp_exponent}
\end{table}

\addBHFigure{3}{2}{3}{2}

\subsection{The action as a function of the ratio of radii}

In  figure \ref{fig:schematic} we drew a schematic picture of $I(\beta)$, the action as a function of the ratio. We expect $I(\beta=0), I(\beta=\infty)$ to diverge.
We also expect the two topologies to approach $\beta=\beta_s$ with zero-slope decayed oscillations. In all cases, we found numerically that the action decreases at second order in the expansion around the singular solution. We haven't found an analytical explanation for this fact. Thus the simplest scenario (like the one drawn in figure \ref{fig:schematic}) would be a single first order phase transition between the two topologies, close to $\beta_s$. 
Figure \ref{fig:S3S2_Fb_1} agrees with this description and exhibits one first order phase transition. The critical ratio of the phase transition for this case agree with the results previously found in~\cite{Blackman:2011in}.

There are two cases that display a somewhat different behavior than that in figures \ref{fig:schematic} and \ref{fig:S3S2_Fb_1}.
The first is the case of equal spheres $d_1=d_2$. The behavior there is very similar besides the sign of the divergence at $\beta \sim 0,\infty$ (we show examples of this in the next section). The second is the Hawking-Page $d_2=1$, shown on the left-hand side of figure \ref{fig:schematic}, where there are no oscillations (as there is no singular solution). 
\addFbFigureSingle{3}{2}{3}{0.720}

\subsection{The stress energy tensor expectation value as a function of \texorpdfstring{$\beta$}{}}

As we mentioned above (see section \ref{sec:sing_sol_pert}) $\gd$ is proportional to the traceless part of the CFT stress tensor expectation value. For $\sd$ we have $\gd^{1,2}= \frac{d^{d}}{d \rho^d}\( r^2_{1,2}(\rho)\) \Big|_{\rho=0}$. Because $\gd$ is traceless the two components are dependent, and it is enough to plot $\gd^{1}$.
Based on the analysis above, we expect $\gd^{1,2}(\beta)$ around $\beta_s$ to have spiral decayed oscillations toward the singular value $\gd^{1,2}(\beta_s)=0$. In figure \ref{fig:S3S2_gd1b_1} we can see exactly this behavior. Similar behavior can be found also for other $d_1,d_2$ (see section \ref{sec:num}). All the graphs display a transition between monotonic behavior at extreme ratios ($\beta \sim 0,\infty$), and the spiral decay around $\beta_s$.
\addgdbFigure{3}{2}{3}{2}

\section{Numerical results for other products of spheres} \label{sec:num}
\subsection{Two equal spheres \texorpdfstring{$d_1=d_2$}{}}

Figures 
\ref{fig:S2S2_bh_1},
\ref{fig:S2S2_Fb_1},
\ref{fig:S2S2_gd1b_1},
\ref{fig:S3S3_bh_1},
\ref{fig:S3S3_Fb_1} and
\ref{fig:S3S3_gd1b_1} describe the  behavior of CFTs on the spaces $S^2\times S^2$ and $S^3 \times S^3$. These cases have a similar qualitative description as in section \ref{sec:num_results}. The only difference is the sign of the coefficient of $\beta^{d_1}$ in the action for $\beta \sim 0,\infty$.
As the two limits are identical, the sign is symmetric, but it depends on $d$. For $d_1=d_2=2$ it is negative (see figure \ref{fig:S2S2_Fb_1}), and for $d_1=d_2=3$ it is positive (see figure \ref{fig:S3S3_Fb_1}).

\addBHFigure{2}{2}{1}{3}
\addFbFigure{2}{2}{1}{3}{1}
\addgdbFigure{2}{2}{3}{2}
\addBHFigure{3}{3}{1}{4}
\addFbFigure{3}{3}{1}{4}{1}
\addgdbFigure{3}{3}{1}{3}

\subsection{The gravity dual of \texorpdfstring{$S^4\times S^2$}{}}

Figures 
\ref{fig:S4S2_bh_1},
\ref{fig:S4S2_Fb_1} and 
\ref{fig:S4S2_gd1b_1} give the ratio of radii, the action and the stress tensor of the $S^4\times S^2$ theory, respectively. The qualitative discussion in section \ref{sec:num_results} is relevant to this case as well. Our results for the phase structure in this case agree with the results previously found in~\cite{Blackman:2011in}.
\addBHFigure{4}{2}{5}{4}
\addFbFigure{4}{2}{5}{6}{0.584}
\addgdbFigure{4}{2}{3}{4}

\section*{Acknowledgements}
We would like to thank M. Berkooz for many useful discussions, and M. Van Raamsdonk for pointing us to the results of~\cite{Blackman:2011in}. This work was supported in part  by an Israel Science Foundation center for excellence grant (grant number 1989/14) and by the Minerva foundation with funding from the Federal German Ministry for Education and Research. OA is the Samuel Sebba Professorial Chair of Pure and Applied Physics. 


\bibliographystyle{JHEP}
\bibliography{refs}

\end{document}